\documentclass[journal=jacsat,manuscript=article]{achemso}

\usepackage[version=3]{mhchem}
\usepackage{amsmath,amssymb}
\usepackage[dvipsnames]{xcolor}
\usepackage{soul}

\author{Kirill A. Bronnikov}
\affiliation{School of Physics and Engineering, ITMO University, St. Petersburg, Russia}
\email{kirill.bronnikov@metalab.ifmo.ru}

\author{Mikhail F. Limonov}
\affiliation{School of Physics and Engineering, ITMO University, St. Petersburg, Russia}
\alsoaffiliation{Ioffe Institute, St. Petersburg, Russia}

\author{Nikolay S. Solodovchenko}
\affiliation{School of Physics and Engineering, ITMO University, St. Petersburg, Russia}

\title{Hot spot as hallmark of transition from dielectric disk to ring}

\abbreviations{DR,RR,WGM,RGM}

\begin{document}

\begin{abstract}
  Topological transformations of dielectric structures radically change the eigenvalues and eigenfunctions of photonic resonances. Moreover, optical effects may arise that characterize the moment of transition from one structure to another, but are not inherent in either the initial or final structure. We demonstrate that such a hallmark of the disk--ring transition is a hot spot of a special nature that arises at the moment a central hole appears in the disk. The hot spot in the air hole is caused by the Mie resonance of the disk with azimuthal number $m$ = 1, while other Mie resonances do not contribute to the effect due to symmetry. As the hole increases, the hot spot fades out, and we theoretically and experimentally observe the formation of photonic resonances of the ring from resonances of the disk. Using near-field and far-field measurements, we discovered clustering of disk photonic modes into distinct galleries of ring eigenmodes that are formed by the inner and outer walls of the cavity. Thus, we demonstrate both the beginning and the end of the rearrangement of photonic eigenmodes during the transition from a dielectric disk to a narrow ring.
\end{abstract}

\section{Introduction}
Topology can be considered a link between bulk and surface properties and therefore plays an important role in the formation of photonic eigenmodes, which generally depend on the resonator size and boundary conditions \cite{luTopologicalStatesPhotonic2016}. The study of the topological properties of non-Hermitian systems has led to the discovery of new photonic states, exemplified by "Fermi arcs", which in photonic structures connect topological degeneracies, with surface Fermi arcs connecting ideal Weyl points, and bulk Fermi arcs connecting exceptional points \cite{koziiNonHermitianTopologicalTheory2017, zhouObservationBulkFermi2018, yangIdealWeylPoints2018}. Recently, topological transitions in various materials have been actively studied, including topological quantum phase transitions, which radically change their photonic and electronic properties \cite{bernevigQuantumSpinHall2006, krishnamoorthyTopologicalTransitionsMetamaterials2012, khanikaevTwodimensionalTopologicalPhotonics2017, zengTailoringTopologicalTransitions2022, sinhaBerryCurvatureDipole2022}. This new direction in physics combines studies of structures in which, when parameters change, transitions occur between different phases that have different topological properties. It all started with the study of semiconductor quantum wells mercury telluride – cadmium telluride. The authors found that with a change in the thickness of the quantum well, a topological quantum phase transition occurs between a conventional insulating phase and a phase exhibiting a quantum spin Hall effect \cite{bernevigQuantumSpinHall2006}. In studies of moiré systems such as twisted bilayer graphene, it was discovered that the Berry curvature dipole exhibits topological transitions in the bands and changes its sign \cite{sinhaBerryCurvatureDipole2022}.

Expanding the horizon of research on topological transitions, in this work we study the change in photonic properties during the transition between two different dielectric structures with different topologies, namely between a disk resonator (DR) and a ring resonator (RR) with a rectangular cross section. This approach is fundamentally different from the studies presented in works \cite{bernevigQuantumSpinHall2006, krishnamoorthyTopologicalTransitionsMetamaterials2012, khanikaevTwodimensionalTopologicalPhotonics2017, zengTailoringTopologicalTransitions2022, sinhaBerryCurvatureDipole2022}, where one object with variable structural or dielectric parameters was studied. We are interested in both the transformation of photonic eigenmodes and changes in the resonant scattering spectra during the DR $\rightarrow$ RR transition. Note that extensive literature has been devoted to the study of the photonic properties of DRs and RRs, which is associated with a wide range of practical applications of these structures, such as filters and switches, antenna elements, modulators, sensors and building blocks of advanced integrated optical circuits \cite{vahalaOpticalMicrocavities2003, bogaertsSiliconMicroringResonators2012, liUsingBackscatteringBackcoupling2019, marpaungIntegratedMicrowavePhotonics2019, goswamiReviewAllopticalLogic2021, goedeModesplittingMicroringResonator2021}. Note that dielectric RRs with a rectangular cross-section have a unique photonic structure in the low-frequency region of the spectrum \cite{solodovchenkoCascadesFanoResonances2022, chetverikovaOpticalFingerprintsDielectric2023a, chetverikovaRadialAxialPhotonic2023}. The scattering spectrum consists of separate galleries, each of which begins with a broad band of radial or axial Fabry-Pérot resonance and continues with an equidistant sequence of narrow azimuthal resonances with exponentially increasing $Q$-factors \cite{solodovchenkoCascadesFanoResonances2022}.

In this work, the emphasis is on experimental and theoretical studies of the transformation of the resonant scattering spectra and photonic eigenmodes with a gradual increase of the coaxial hole in the DR to the ratio of the inner and outer radii of the RR $R_{in}/R_{out} = 0.6$. Among the original results, we highlight the detection of a hot spot \cite{bakkerMagneticElectricHotspots2015} almost immediately with the appearance of an air hole in the DR ($R_{in}/R_{out} = 5 \times 10^{-5}$), which fades when the hole increases to $R_{in}/R_{out} \sim 0.1$ and higher (permittivity of the resonator $\varepsilon$ = 43). Thus, a hot spot can be considered a hallmark of a DR $\rightarrow$ RR transition. We established that this hot spot is caused by the Mie resonance with specific symmetry (azimuthal number $m$ = 1), which determines the intense scattering. Next, using the experimental technique of far- and near-field spectroscopy, we studied in detail the formation of ring gallery modes, which, from a photonics point of view, completes the transition between two structures with different topologies.

This work is a key step in our research into the transformation of the photonic properties of dielectric resonators in a chain of transitions between structures with different topologies and curvatures: disk – ring – split ring – cuboid \cite{bochkarevSplitRingObius2023}.

\section{Methods}

\subsection{Sample fabrication}

The experimental samples were two ceramic disks with a radius $R_{out}$ of 25 mm and height $h$ of 3 mm made of the MgO-CaO-TiO$_2$ and LaAlO$_3$-CaTiO$_3$ compounds with different dielectric permittivities $\varepsilon \approx 20$ and 45, loss tangent tan$\delta \approx 0.8 \times 10^{-4}$ and $1 \times 10^{-4}$, respectively. The actual permittivity values of the disk samples were determined by measuring the SCS spectra and fitting them with the calculated ones, which resulted in the following adjusted parameters: $\varepsilon = 19.32$ and 43.85. The variation of the $R_{in}/R_{out}$ parameter was performed by systematic drilling and enlargement of the hole in the samples, while measuring the total scattering cross-section (SCS) spectrum for each $R_{in}/R_{out}$ value with a step of 0.01.

\subsection{Experiment}

The SCS spectra were measured in the microwave spectral range in an anechoic chamber. Samples of the resonators were placed in the middle between two wideband horn antennas (Trim TMA 1.0--18.0 GHz) separated by a distance of $\sim$4 m. One of the antennas acted as an emitter providing near-plane-wave excitation, and the other was a receiver. Both antennas were connected to the Rohde \& Schwarz ZVB 20 vector analyzer with the working range of 0.01--20 GHz. In the experiment, the frequency range of 1--20 GHz was swept with 32001 data points. The SCS was then obtained using the optical theorem as follows:
\[\sigma_{ext} = -\frac{4 \pi c}{\omega}~\text{Im} \left( \frac{S_{21}}{S_{21}^{free}} \right), \]
where $S_{21}$ and $S_{21}^{free}$ are complex measured transmission coefficients between two antennas in the presence of a sample and in free space, respectively. 

The near-field study was performed by measuring the $z$-component of the magnetic field $H_z$ using the LANGER EMV-Technik SX near field probe. The probe was placed approximately 5 mm above the surface of the sample and scanned in the horizontal plane line-by-line.

$Q$-factor was obtained by fitting the resonant peaks from experimental SCS spectra with the Fano formula:
\[\sigma_{SCS} = a \frac{(q + \Omega)^2}{1 + \Omega^2} + p, \quad \Omega = \frac{x - x_0}{\Gamma/2}, \]
where $q$ is the Fano parameter, $a$ is the amplitude and $p$ is an additional fitting parameter describing the background, $x_0$ is the resonant size parameter (resonant frequency), and $\Gamma$ is the width of the resonant line. The $Q$-factor is then obtained as $Q = x_0/\Gamma$.

\subsection{Numerical modeling}

Calculations of the eigenfrequencies, SCS spectra, and near-field distributions were performed using COMSOL Multiphysics software controlled by custom Python code. In the solved model a plane wave was incident on the isolated dielectric DR/RR with material and geometric parameters matching those of the experimental samples. The electric field polarization and wavevector were in the plane of the resonator. During the calculations, the $R_{in}/R_{out}$ parameter was varied similarly to the experiment. To increase the speed of calculations, the SCS map was calculated using the azimuthal harmonics with different number $m$, which allows solving a 2D $(\rho,z)$ problem with axial symmetry instead of a 3D $(x,y,z)$ one. In the considered low-frequency region, the scattering of a plane wave is completely described by azimuthal harmonics with $m \leq 25$. The contribution of higher azimuthal harmonics is practically equal to zero over the entire frequency range of interest. More details on the modeling methods can be found elsewhere \cite{chetverikovaRadialAxialPhotonic2023}.

\section{Hot spot: appearance and disappearance}

The transition from a dielectric disk to a ring occurs with the appearance of an internal hole. One of the striking effects of such a transition between two structures with different topologies is the formation of a hot spot in the hole for the electric dipole mode in TE polarization ($m$ = 1 azimuthal harmonic). This effect can be explained using the three-layer Mie theory for an infinite cylinder with a coaxial hole \cite{huangOptimizationPhotonicNanojets2019}. In the case of TE polarization, the electric field inside the hole is represented using the following equations:
\begin{equation}
E_\rho = -\frac{a_m m}{k_0 \rho} J_m(k_0 \rho) e^{im\phi},
\label{eq:1}
\end{equation}
\begin{equation}
E_\phi = a_m \frac{J_{m-1}(k_0 \rho) - J_{m+1}(k_0 \rho)}{2 i} e^{im\phi},
\label{eq:2}
\end{equation}
where $E_\rho$ and $E_\phi$ are the radial and angular components of the electric field, respectively, $k_0 = \omega/c$ is the vacuum wave number, $a_m$ and $J_m (k_0 \rho)$ are the Lorenz-Mie coefficient and Bessel function of order $m$, respectively.

\begin{figure}
\includegraphics[width=0.82\textwidth]{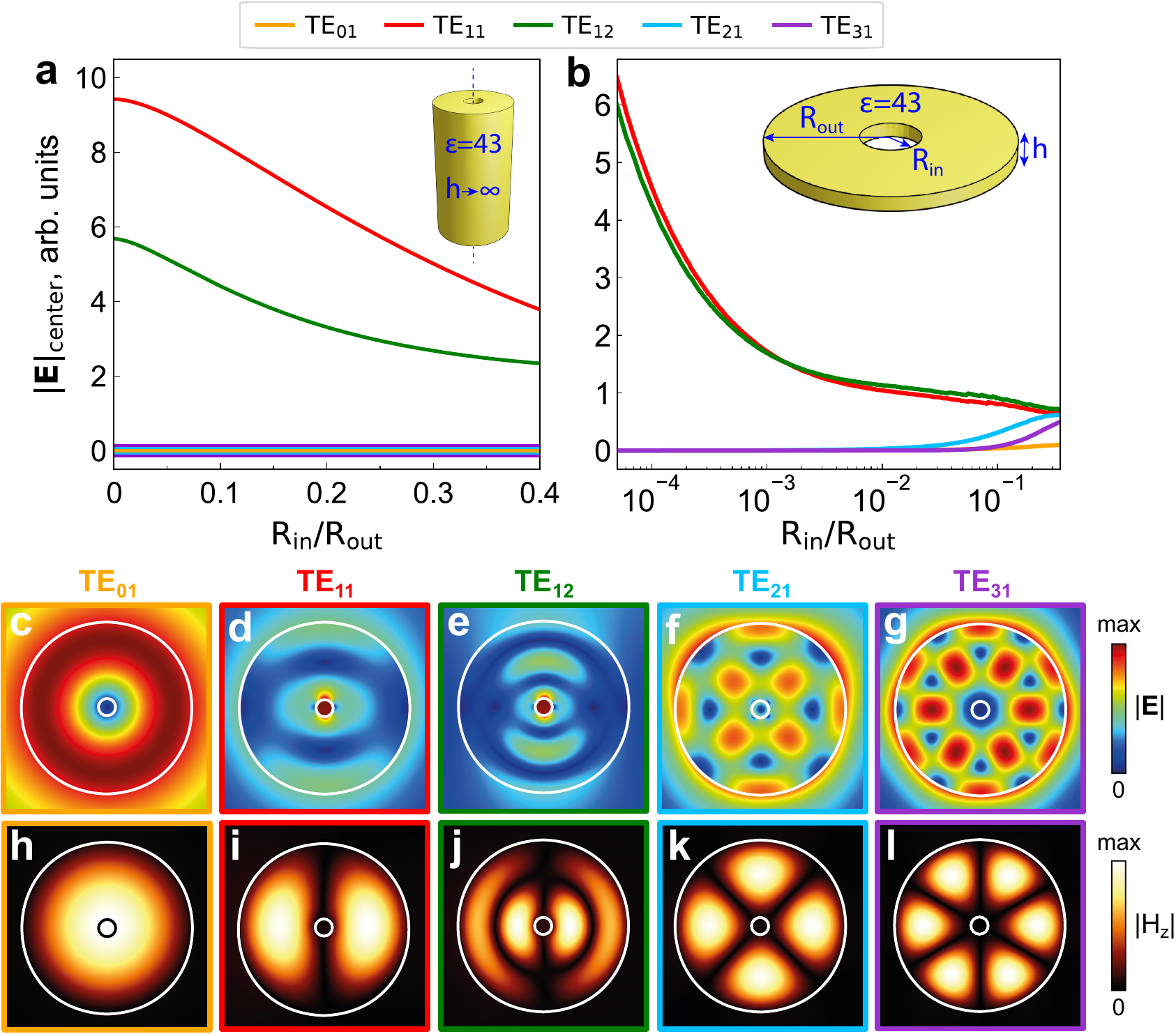}
\caption{Appearance of the electric hot spot in the air hole of the RR. (a, b) Absolute value of the electric field in the center of the resonator for different modes TE$_{mr}$ in the case of (a) infinite RR (2D analytical Mie solution), and (b) finite RR (height-to-radius ratio $h/R_{out} = 0.12$) used in the experiment (3D COMSOL calculation). Insets in (a) and (b) illustrate the geometry of the resonators. (c-g) Electric and (h-l) magnetic field distributions for corresponding modes calculated using the 2D analytical Mie solution for $R_{in}/R_{out}$ = 0.1 (note the maximum of the electric field in the central hole for modes TE$_{11}$ and TE$_{12}$). The permittivity of the RR is $\varepsilon = 43$ in all calculations.}
\label{fig:fig1}
\end{figure}

By sequentially examining each azimuthal harmonic with the electric field components defined by Eq.~(\ref{eq:1}) and (\ref{eq:2}), one can be convinced that the field maximum in the center of the resonator ($\rho$ = 0) exists only for $m$ = 1. Taking into account that all Bessel functions of the order higher than 0 are zero at the origin we have the following cases:

{\setstretch{1}
    \begin{alignat*}{2}
        &m = 0: \quad && E_\rho = 0, E_\phi \sim J_1 (k_0 \rho) = 0, \\[.2cm]
        &m = 1: \quad && E_\rho = -a_1 \dfrac{J_1 (k_0 \rho)}{k_0 \rho} e^{i\phi} \xrightarrow {\rho \rightarrow 0} -\dfrac{a_1}{2} e^{i\phi}, \\[.2cm]
        & \quad && E_\phi = a_1 \dfrac{J_0 (k_0 \rho) - J_2 (k_0 \rho)}{2i} e^{i\phi} = \dfrac{a_1}{2} e^{i(\phi - \pi/2)}, \\[.2cm]
        &m \geq 2: \quad && E_\rho = E_\phi = 0.
    \end{alignat*}
}

Fig.~\ref{fig:fig1}(a) shows the variation of the total electric field amplitude $|E|$ in the center of the RR for the first two modes with the azimuthal index $m$ = 1 and for other modes with $m \neq 1$ when the ratio $R_{in}/R_{out}$ is changed. In this case, we denote the modes as TE$_{mr}$, where index $m$ has the same meaning and $r$ is the radial index, which simultaneously corresponds to the resonance number in the 2D Mie problem. As seen from the electric field distributions provided in Fig.~\ref{fig:fig1}(c-g) for the ratio $R_{in}/R_{out}$ = 0.1, the maximum at the center appears only for modes TE$_{11}$ (Fig.~\ref{fig:fig1}(d)) and TE$_{12}$ (Fig.~\ref{fig:fig1}(e)), whereas other azimuthal harmonics are practically zero in the hole. Magnetic field distributions, in turn, do not demonstrate such striking features (see Fig.~\ref{fig:fig1}(h-l)).

This theoretical approach helps explain the peculiarities of the disk-ring transition in the simple 2D case, which is equivalent to an infinite cylinder/ring in 3D. To address the case of a finite resonator in 3D geometry we investigate the transformation of photonic resonances by gradually increasing the internal opening in the disk, both numerically and experimentally.

Similarly to the analysis with the 2D Mie theory presented above, we calculated the electric field amplitude in the center of the disk/ring with the height-to-radius ratio $h/R_{out} = 0.12$ depending on the $R_{in}/R_{out}$ parameter using COMSOL Multiphysics (Fig.~\ref{fig:fig1}(b)). In the case of a finite-height RR the strength of the electric hot spot for TE$_{11}$ modes fades out significantly faster with $R_{in}/R_{out}$ in comparison to the infinite resonator (note the logarithmic $R_{in}/R_{out}$ scale in Fig.~\ref{fig:fig1}(b)). In addition, fields of other modes with $m \neq 1$ start increasing at $R_{in}/R_{out} \sim 0.01$ and become comparable to the field amplitude of the TE$_{11}$ mode at $R_{in}/R_{out} \sim 0.1$.

\begin{figure}
\includegraphics[width=\textwidth]{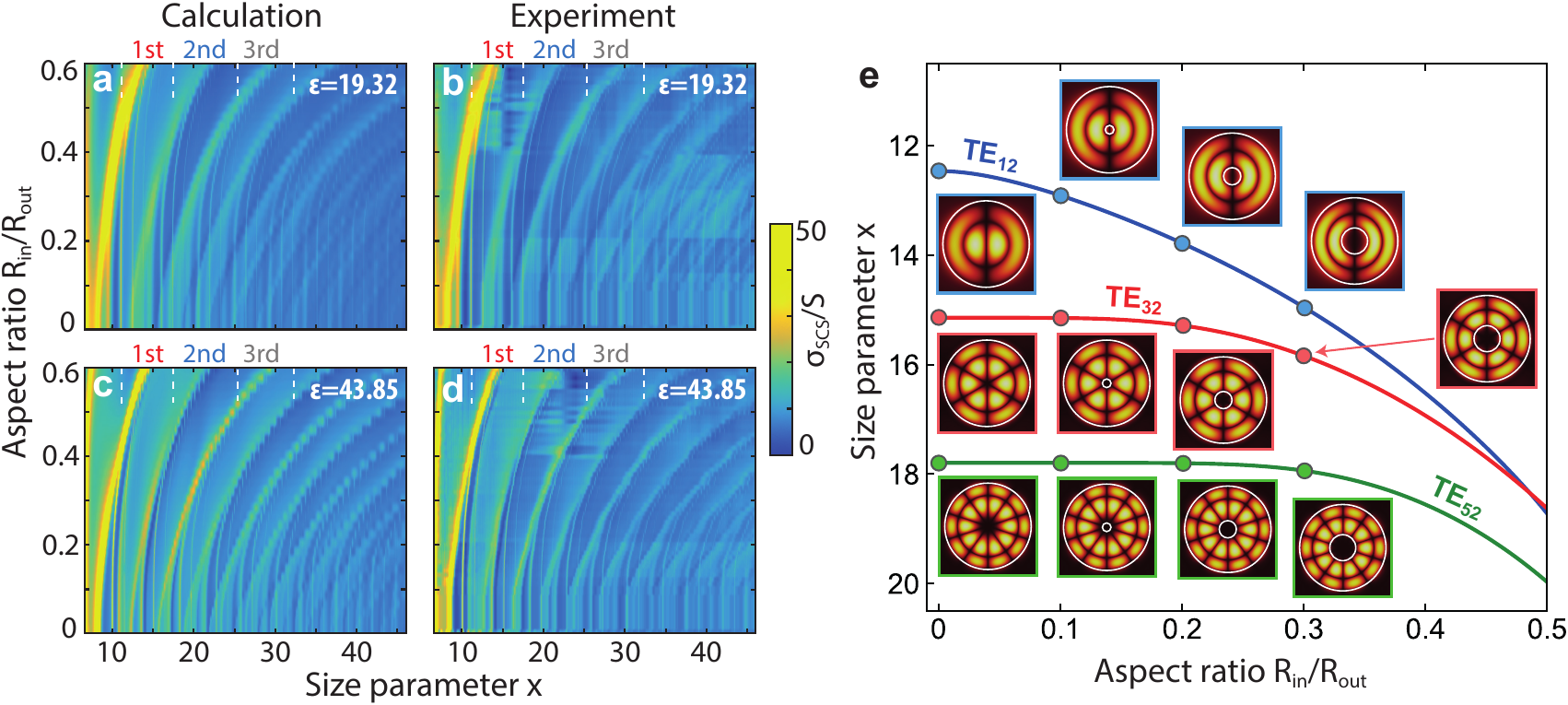}
\caption{(a-d) Calculated and experimental maps of the total $SCS$ (normalized to the geometrical cross-section $S = 2R_{out}h$ of the ring) vs normalized size parameter $x = \omega/c \sqrt{\varepsilon} R_{out}$ and geometrical aspect ratio $R_{in}/R_{out}$ for different dielectric permittivity of the ring. $\varepsilon$ = 19.32: calculation (a), experiment (b); $\varepsilon$ = 43.85: calculation (c), experiment (d). First three ring galleries of modes (1st, 2nd, 3rd) at $R_{in}/R_{out}$ = 0.6 are indicated by white dashed lines and corresponding labels. (e) Dependency of calculated eigenfrequencies of modes with azimuthal indices $m$ = 1, 3, 5 from the 2nd radial gallery on the aspect ratio $R_{in}/R_{out}$ for the ring resonator with $\varepsilon$ = 43.85. Calculated distributions of $|H_z|$ inside the ring are provided in the insets at corresponding parameters indicated by the circles. The height-to-radius ratio is $h/R_{out} = 0.12$ in all experiments and calculation.}
\label{fig:fig2}
\end{figure}

\section{Formation of photonic galleries of dielectric ring}

The second part of the work was to study changes in the RR scattering spectra with gradual narrowing of the ring (increasing $R_{in}/R_{out}$) in order to determine the parameters at which photonic eigenmodes of the RR are formed. Fig.~\ref{fig:fig2}(a-d) shows the calculated and experimentally measured total SCS spectra depending on the size parameter $x = \omega/c \sqrt{\varepsilon} R_{out}$ and ratio $R_{in}/R_{out}$ for dielectric RRs with different permittivities $\varepsilon$ = 19.32 and 43.85. As can be seen, the calculated spectra closely follow the experimentally obtained dependency with gradual transformation of the equidistant set of whispering gallery modes (WGMs) of a DR ($R_{in}/R_{out}$ = 0) to the spectra of a ring with groups of resonances gathered according to the radial indices of the modes. Each of the groups begins with a broad Fabry-P{\'e}rot-type resonance formed by quantization between the inner and outer walls of the ring (radial index $r$ = 1, 2, 3, ...), followed by equidistant longitudinal modes with increasing azimuthal index $m$. These groups, by analogy with a disk resonator, were called ring gallery modes (RGMs) \cite{solodovchenkoCascadesFanoResonances2022}.

It can be noted that in addition to grouping of the resonances into separate RGMs, each mode of the ring gallery has a different frequency shift between the initial position in the disk spectrum and the final position in the ring spectrum. We explain this behavior by analyzing the field distribution of the resonant modes inside the resonator. As an example, we consider the dependency of the eigenfrequencies of three longitudinal modes ($m$ = 1, 3, 5) from the second gallery ($r$ = 2) on the $R_{in}/R_{out}$ ratio in Fig.~\ref{fig:fig2}(e) together with the corresponding calculated field profiles provided as insets. If the resonant mode of the DR has a field maximum located close to the center of the disk, as in the case of the $m$ = 1 mode, then when the coaxial hole expands, the mode field distribution will be adjusted according to the new internal boundary, and the frequency increases due to narrowing of the resonator in the radial direction. In another case, if there is a zero field at the center of the disk, as with $m$ = 3 or 5, then the resonant frequency will not change significantly until the inner wall of the ring reaches the area of the non-zero mode field. For example, for the $m$ = 5 mode, the frequency begins to increase rapidly at $R_{in}/R_{out}$ = 0.3 (Fig.~\ref{fig:fig2}(e), green line).

\begin{figure}
\includegraphics[width=\textwidth]{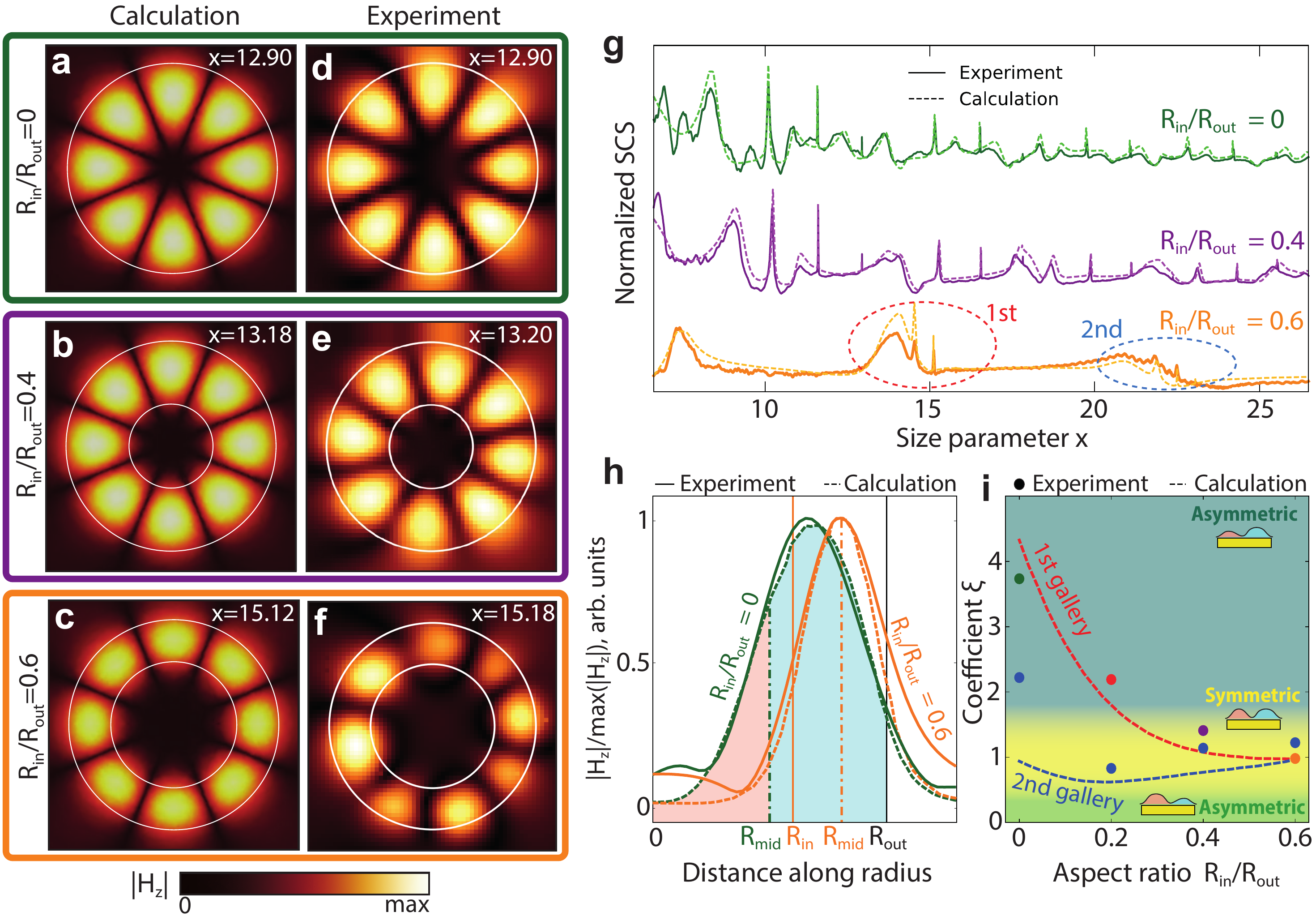}
\caption{Near-field study of plane wave scattering from the ring resonator with $\varepsilon$ = 43.85. Calculated (a-c) and measured (d-f) near-field distributions of $|H_z|$ for the $(m, r)$ = (4, 1) mode at the ratio $R_{in}/R_{out}$ = 0 (a, d), 0.4 (b, e), 0.6 (c, f). (g) Total calculated and measured SCS spectra of rings with corresponding $R_{in}/R_{out}$ ratios; the 1st and 2nd ring galleries at $R_{in}/R_{out}$ = 0.6 are indicated by dotted ellipses. (h) Experimental and calculated distributions of the normalized $|H_z|$ along the radial direction for the $(m, r)$ = (4, 1) mode at $R_{in}/R_{out}$ = 0 (green curves) and $R_{in}/R_{out}$ = 0.6 (orange curves). (i) Dependence of the symmetry parameter $\xi$ on the ratio $R_{in}/R_{out}$ derived from the experimental and calculated field distributions of modes $(m,r)$ = (4, 1) (1st gallery) and (5, 2) (2nd gallery).}
\label{fig:fig4}
\end{figure}

Next, we performed a near-field study of the photonic resonances. A comparison of the calculated and experimental distributions of the magnetic field $|H_z|$ of the mode ($m$, $r$) = (4, 1) in the $XY$ plane (orthogonal to the ring axis) is presented in Fig.~\ref{fig:fig4}(a-f) for different values of $R_{in}/R_{out}$ = 0, 0.4, and 0.6. The appearance and enlargement of the hole leads to a change in the field distribution and a decrease of the area occupied by the mode field. Furthermore, the width of the ring decreases with increasing of $R_{in}/R_{out}$, which leads to a spectral shift of resonances. The corresponding SCS spectra for $R_{in}/R_{out}$ = 0, 0.4, and 0.6 are shown in Fig.~\ref{fig:fig4}(g).

While appearance of the hot spot in the air hole can be considered as an indicator of the start of DR $\rightarrow$ RR transition, the criterion of the end of this transition should be introduced. For this, we consider the mode field distribution along the radial direction of the ring. In the case of a disk and a ring with small $R_{in}/R_{out}$, the mode field is rather asymmetric relative to the central line $R_{mid} = (R_{in} + R_{out})/2$ and concentrates closer to the outer wall of the resonator, which is typical for WGMs. With further increase of $R_{in}/R_{out}$ the mode field is tightened between the inner and outer boundaries and becomes more symmetric relative to $R_{mid}$ (Fig.~\ref{fig:fig4}(h)). Due to the difference between field patterns of different modes, each of them will reach such symmetrical distribution at their own value of $R_{in}/R_{out}$.

In order to quantitatively assess this process, we introduce the mode symmetry parameter
\[
    \xi = \frac{\int_{R_{mid}}^{R_{out}} |H_z(\rho)|\,d\rho}{\int_{R_{in}}^{R_{mid}} |H_z(\rho)|\,d\rho},
\]
where the field amplitude is integrated along the radial coordinate $\rho$ over the inner (denominator) and outer (numerator) half of the ring width. The $z$ component of the magnetic field is chosen here to compare the numerical modeling with experiment where $H_z$ was measured. As an example, Fig.~\ref{fig:fig4}(h) shows the experimental and calculated normalized magnetic field distributions along the radius for the disk (green color) and the ring with $R_{in}/R_{out}$ = 0.6 (orange color). In the case of DR ($R_{in}/R_{out}$ = 0), the intervals of integration are indicated by pink (inner half of the ring width) and cyan (outer half) shaded areas. The parameter $\xi$ was calculated from the modeled and experimental field patterns of the modes $(m,r)$ = (4, 1) and (5, 2), and its dependence on the ratio $R_{in}/R_{out}$ is plotted in Fig.~\ref{fig:fig4}(i). As one can see, $\xi > 1$ for the DR and RR with small $R_{in}/R_{out}$ indicating high asymmetry of the field distribution. With increasing hole size $\xi$ approaches 1 and $|H_z|$ becomes almost symmetric, which corresponds to the RR modes \cite{solodovchenkoCascadesFanoResonances2022}. The value of $R_{in}/R_{out}$ at which $\xi$ stabilizes at $\approx$1 can be then considered as the end of the transition from WGMs of the disk to RGMs of the ring.

\begin{figure}
\includegraphics[width=0.65\textwidth]{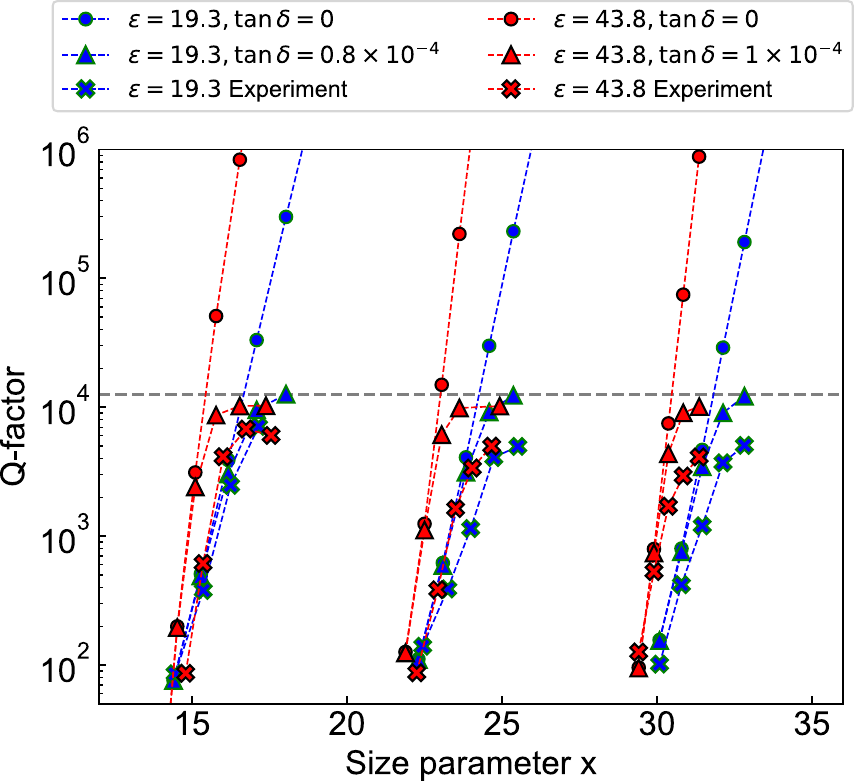}
\caption{Calculated and experimental $Q$-factors of modes of the first three galleries of rings with $R_{in}/R_{out}$ = 0.6 and different permittivity $\varepsilon$. Circles and triangles represent calculation for lossless rings and rings with loss tangent typical for ceramic samples used in the experiment, respectively. Crosses represent experimental data. The horizontal gray dashed line indicates the maximum achievable $Q$-factor value $Q_{max} = 1/\text{tan}\delta$ with tan$\delta = 0.8 \times 10^{-4}$ for the ring with $\varepsilon$ = 19.32. Dashed lines connecting the data points are a guide for the eye.}
\label{fig:fig6}
\end{figure}

The $Q$-factors obtained from the calculated and experimental SCS spectra for the resonant modes from the first three galleries for the rings with $R_{in}/R_{out}$ = 0.6 and different permittivities $\varepsilon$ are presented in Fig.~\ref{fig:fig6}. In the case of lossless ring resonators, the $Q$-factors of the modes in each gallery follow nearly linear dependencies on the resonant frequency in a logarithmic scale (circles in Fig.~\ref{fig:fig6}). Adding absorption to the material leads to “saturation” of the $Q$-factor at a maximum value defined by $Q_{max} = 1/\text{tan}\delta$, where tan$\delta$ is the loss tangent (triangles in Fig.~\ref{fig:fig6}).

These calculations qualitatively agree with the experimental results indicated by the crosses in Fig.~\ref{fig:fig6}. The observed deviations in the $Q$-factor can be explained by the actual loss tangent values of the ceramic ring samples being different from those provided in the datasheet. The mismatch in the spectral positions of the resonances is due to the limited accuracy of the inner hole fabrication, which results in slightly different values of $R_{in}/R_{out}$ and dispersion of dielectric losses, which increase with increasing frequency.

Asymmetric modes with large azimuthal number $m$ have a significant quality factor $(Q_{rad}>10^8)$, which excludes the possibility of their detection in scattering spectra. Since any resonator has non-zero material losses, when the radiation $Q_{rad}$-factor exceeds the level of the material $Q_{mat}$-factor, the amplitude of the resonance decreases. Therefore, only a few high-$Q$ resonances are visible in the scattering spectra in each gallery.

\section{Conclusion}

The notable photonic effect --- the formation of a hot spot --- signalizes the transition between two structures with different topologies that occurs once a coaxial hole is formed in a dielectric disk and it becomes a ring. This hot spot appears only for Mie modes of the DR, which have an electric field maximum at the center of the resonator, and these are modes with azimuthal number $m$ = 1, which readily follows from the Mie solutions. In contrast to the infinite RR approximation, the hot spot intensity quickly decreases with increasing hole radius. Hot spots with light concentrated in subwavelength regions are widely used for biosensing, photocatalysis, nonlinear light generation, and other applications \cite{mohammadiAccessibleSuperchiralNearFields2019,caldarolaNonplasmonicNanoantennasSurface2015,kuznetsovOpticallyResonantDielectric2016}.

The field distribution of each mode also defines the dynamics of the transformation of the scattering properties of the RR upon increasing the coaxial hole radius. Namely, resonances in the ring scattering spectra start shifting to higher frequencies when the hole boundary reaches the area of the non-zero field of the respective mode. This was demonstrated using the example of the calculated field patterns of various modes with different azimuthal indices ($m$ = 1, 3, 5). As a result, each mode experiences a different spectral shift, and we observe the grouping of resonances in the scattering spectrum of the RR according to their radial indices with the formation of photonic ring galleries of modes both in calculation and experiment. Measurements of the near-field patterns allowed us to verify these findings. 

As the mode field patterns inside the RR are almost symmetrical relative to the middle line of the ring, the size of the hole at which the modes reach such a symmetric state can be considered as the end of the DR $\rightarrow$ RR transition. We demonstrate this by introducing the symmetry parameter $\xi$, which is equal to 1 when the mode field distribution is symmetric and deviates from unity in the case of asymmetrical field pattern. In this approach, modes with small azimuthal index $m$ arrive at the end of the DR $\rightarrow$ RR transition earlier than the modes with large $m$.

It was also demonstrated experimentally that the $Q$-factor in each ring gallery increases almost exponentially for every subsequent azimuthal mode until it reaches the limit $Q_{max}$ determined by the losses in the material. The presented results can be useful in designing novel elements of resonant all-dielectric photonics.

\begin{acknowledgement}
    KB acknowledges the financial support by the ITMO Fellowship Program. ML acknowledges the financial support from the Russian Science Foundation (Project 23-12-00114). NS acknowledges the financial support from the Foundation for the Advancement of Theoretical Physics and Mathematics “BASIS” (Russia).
\end{acknowledgement}

\bibliography{refs}

\end{document}